\documentclass[osajnl,twocolumn,showpacs,preprintnumbers]{revtex4}
\usepackage{epsfig}
\newcommand{\ave}[1]{\ensuremath{\langle #1 \rangle}}
\newcommand{\D}{\Delta}
\newcommand{\de}{\delta}
\begin{document}

\title{\bf Experimental observation of three-color optical quantum correlations}
\author{K. N. Cassemiro, A. S. Villar, P. Valente, M. Martinelli, and P.
Nussenzveig}
\email{nussen@if.usp.br}
\affiliation{Instituto de F\'\i sica, Universidade de S\~ao Paulo,
Caixa Postal 66318, 05315-970 S\~ao Paulo, SP, Brazil} 

\begin{abstract}
Quantum correlations between bright pump, signal, and idler beams produced by an 
optical parametric oscillator, all with different frequencies, are experimentally 
demonstrated. We show that the degree of entanglement between signal and idler 
fields is improved by using information of pump fluctuations. This is the first 
observation of three-color optical quantum correlations. 
\end{abstract}

\vspace{10pt}

\maketitle

Quantum correlations are a signature of nonclassical light generation. 
The optical parametric oscillator (OPO) is the best-known and most 
widely used source of such correlations. Squeezing in the intensity difference 
of the twin beams it produces above-threshold (signal and idler)~\cite{firsttwin87} 
reached the record value of $-9.7$~dB~\cite{fabre10dB}. Bipartite continuous variable 
(CV) entanglement in this system, which requires the observation of phase 
anticorrelations as well, was only demonstrated very 
recently~\cite{prlentangtwinopo,optlettpeng,pfisterentang}. 
The parametric process involves three fields, yet the pump field is typically 
treated as a classical quantity. As an exception, quantum properties of the pump 
were first measured by Kasai {\it et al.}~\cite{kasaipump}. This arises a 
natural question: are there quantum correlations between all three fields? An 
affirmative answer has been recently given by Villar {\it et al.}
~\cite{prltrientangopo}, who investigated the problem theoretically. Here, 
we provide the first experimental affirmative answer by observing triple 
correlations between quadratures of pump, signal, and idler fields.

Beyond the demonstration of nonclassical light features, one should notice that 
all three fields have different frequencies. The interest of quantum frequency 
conversion was noticed in the early 1990s~\cite{kumar1,kumar2} and opens 
perspectives for the interaction of light with physical systems having different resonance 
frequencies. The above-threshold OPO produces, in general, nondegenerate twin 
beams and the pump beam has approximately twice their frequencies. Three-color 
quantum correlations increase the number of physical systems that can be 
simultaneously investigated. Correlations with the pump can also be used to 
enhance the bipartite entanglement between the twin beams, as we show below. 

Quantum correlations should exist between the phase quadrature of the 
pump field and the sum of phase quadratures of signal and idler fields as 
a direct consequence of energy conservation $\omega_0=\omega_1+\omega_2$. 
Indeed, by relating frequency fluctuations to phase fluctuations, we obtain 
$\de\phi_0=\de\phi_1+\de\phi_2$. Indices $j\in\{0,1,2\}$ 
refer to pump, signal and idler fields, respectively. The quadratures are 
defined through the field annihilation operators $\hat a_j=\exp(i\phi_j)(\hat 
p_j+i\hat q_j)$, where $\phi_j$ is chosen so that $\ave{\hat q_j}=0$. When the 
OPO is detuned from exact triple resonance, this phase-phase correlation, 
$C_{\hat q_0\hat q_+}=\ave{\de\hat q_0\,\de\hat q_+}$, is partially transferred to 
an amplitude-phase correlation, $C_{\hat p_0\hat q_+}=\ave{\de\hat p_0\,\de\hat q_+}$, owing 
to phase noise to amplitude noise conversion~\cite{yabuzaki} inside the OPO cavity 
$[\hat q_+ \equiv (\hat q_1 + \hat q_2)/\sqrt{2}]$. Our experiment is designed 
to measure joint fluctuations of a combination of $\hat q_+$ and $\hat p_0$ and 
compare them to the shot noise level, which defines the standard quantum limit (SQL). 
This will enable us to improve the bipartite entanglement of the twin beams. 

Twin beam entanglement is proven by violation of an inequality derived by Duan 
{\it et al.}~\cite{dgcz} and Simon~\cite{simon}. Van Loock and Furusawa~\cite{loockfuru} 
generalized it to include a third field: 
\begin{equation}
\label{ineqtriple}
\D^2\hat p_-+\D^2(\hat q_+-\alpha_0\,\hat p_0)\geq 2\;, \quad \mathrm{with} \quad 
\alpha_0=\frac{C_{\hat p_0\hat q_+}}{\D^2\hat p_0}\;.
\end{equation}
Here $\alpha_0$ is a parameter chosen to minimize the left-hand side of the 
expression above and $\hat p_- \equiv (\hat p_1 + \hat p_2)/\sqrt{2}$. Each term 
$\D^2\hat p_-$ and $\D^2(\hat q_+-\alpha_0\,\hat p_0)$ is normalized to the SQL. Our 
attention will be focused on the corrected phase sum noise (second term above), which 
can be rewritten as
\begin{equation}
\label{ineqtriplecorr}
\D^2\hat q_+' \equiv \D^2\hat q_+-\beta_0\;, \quad \mathrm{with} 
\quad \beta_0=\frac{C_{\hat p_0\hat q_+}^2}{\D^2\hat p_0}. 
\end{equation}
If $\D^2\hat q_+' < 1$ and $\beta_0\ne0$, there is a quantum correlation between $\hat p_0$ and 
$\hat q_+$. 

\begin{figure}[ht]
\centering
\epsfig{file=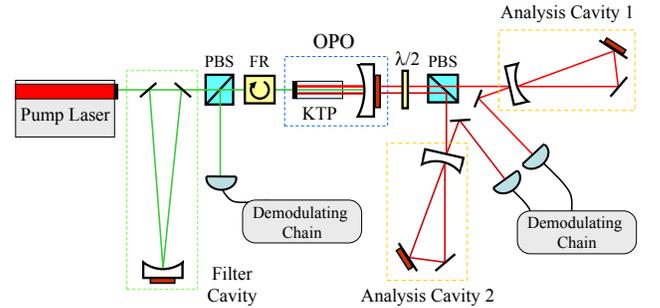,scale=0.39}
\caption{(Color online) Sketch of the experimental setup. PBS: polarizing beam splitter cube; 
FR: Faraday rotator $\lambda/2$: half-wave plate.}
\label{fig1}
\end{figure}

The experimental setup is sketched in Fig.~\ref{fig1}. The triply resonant 
type-II OPO is pumped by a frequency doubled diode-pumped Nd:YAG laser 
(Innolight Diabolo) at 532 nm. This laser is first transmitted through a filter 
cavity (bandwidth of 2.4~MHz) prior to injection in the OPO, which removes all 
classical noise for analysis frequencies above 15~MHz. It is important to have 
a shot noise-limited pump, since excess pump phase noise is converted to excess 
noise in $\hat q_+$, thus hindering twin beam entanglement~\cite{optcomm04}. The 
nonlinear crystal is a 10~mm long Potassium Titanyl Phosphate (KTP) from Litton. 
The OPO cavity input mirror is flat, directly coated on one crystal surface, with 
97\% reflection at 532~nm and high reflectivity (R$>$99.8\%) at 1064~nm. The other 
crystal surface is anti-reflection coated for both wavelengths (R$<$3\% at 532~nm 
and R$<$0.25\% at 1064~nm). The spherical output mirror is a high reflector for 532~nm 
(R$>$99.8\%) and partial reflector for 1064~nm, R=96\%, with a curvature radius of 
25~mm. The OPO cavity bandwidth for 1064~nm is 50~MHz and the threshold power is 12~mW. 
Orthogonally polarized signal and idler beams are separated by a polarizing beam splitter 
(PBS). In order to measure their quadrature noise, each beam is reflected off an 
analysis optical cavity, which converts phase noise to amplitude noise as a function 
of its detuning~\cite{levenson,galatola}. The twin beams are finally detected on high quantum 
efficiency ($>93\%$) photodiodes (Epitaxx ETX300). Both analysis cavities have bandwidths 
of 14(1) MHz. Overall detection efficiency is $\eta=80$\%. Signal and idler optical 
frequencies differ by approximately 0.35 THz, corresponding to $\Delta\lambda= 1.3$~nm 
in wavelength. Photocurrents are recorded as a function of time as both cavities are 
synchronously scanned. At the same time, amplitude fluctuations of the reflected pump 
beam (extracted by means of a PBS and a Faraday Rotator) are recorded by another photodetector 
(EG\&G FND100, quantum efficiency 60\%), with an overall detection efficiency of 
$\eta_0=45$\% (we are currently working to improve this value by using higher quantum 
efficiency photodiodes). Noise power spectra are obtained by direct 
demodulation of the photocurrents. Each photocurrent is electronically mixed with the 
same sinusoidal reference at the analysis frequency $\nu=27$~MHz and the low frequency 
beat signal is sampled at a 600~kHz repetition rate by an analog-to-digital (A/D) 
card connected to a computer. Variances of these fluctuations are then calculated 
by taking groups of 1000 points, and finally normalized to the SQL.

\begin{figure}[ht]
\centering
\epsfig{file=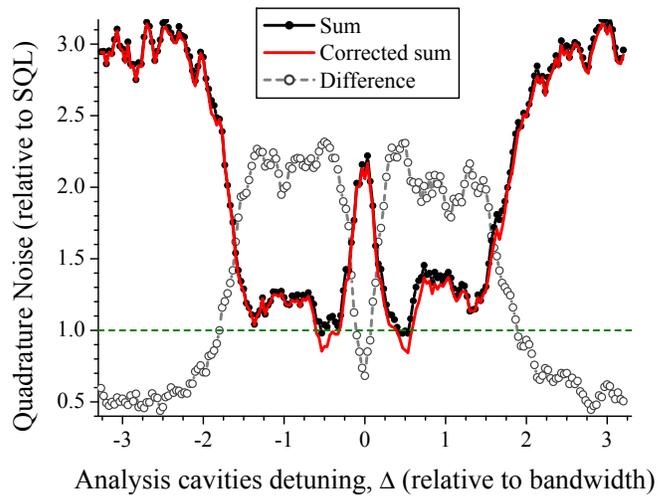,scale=0.35}
\caption{(Color online) Noise spectra at 27MHz, as a function of analysis cavities' 
detuning. Sum of twin beam quadratures: line + full circles; difference 
of twin beam quadratures: line + open circles; sum of twin beam quadratures 
corrected by correlations with the pump amplitude: full line. Shot noise level
is the dashed line. $\sigma=1.34$.}
\label{fig2}
\end{figure}

\begin{figure}[ht]
\centering
\epsfig{file=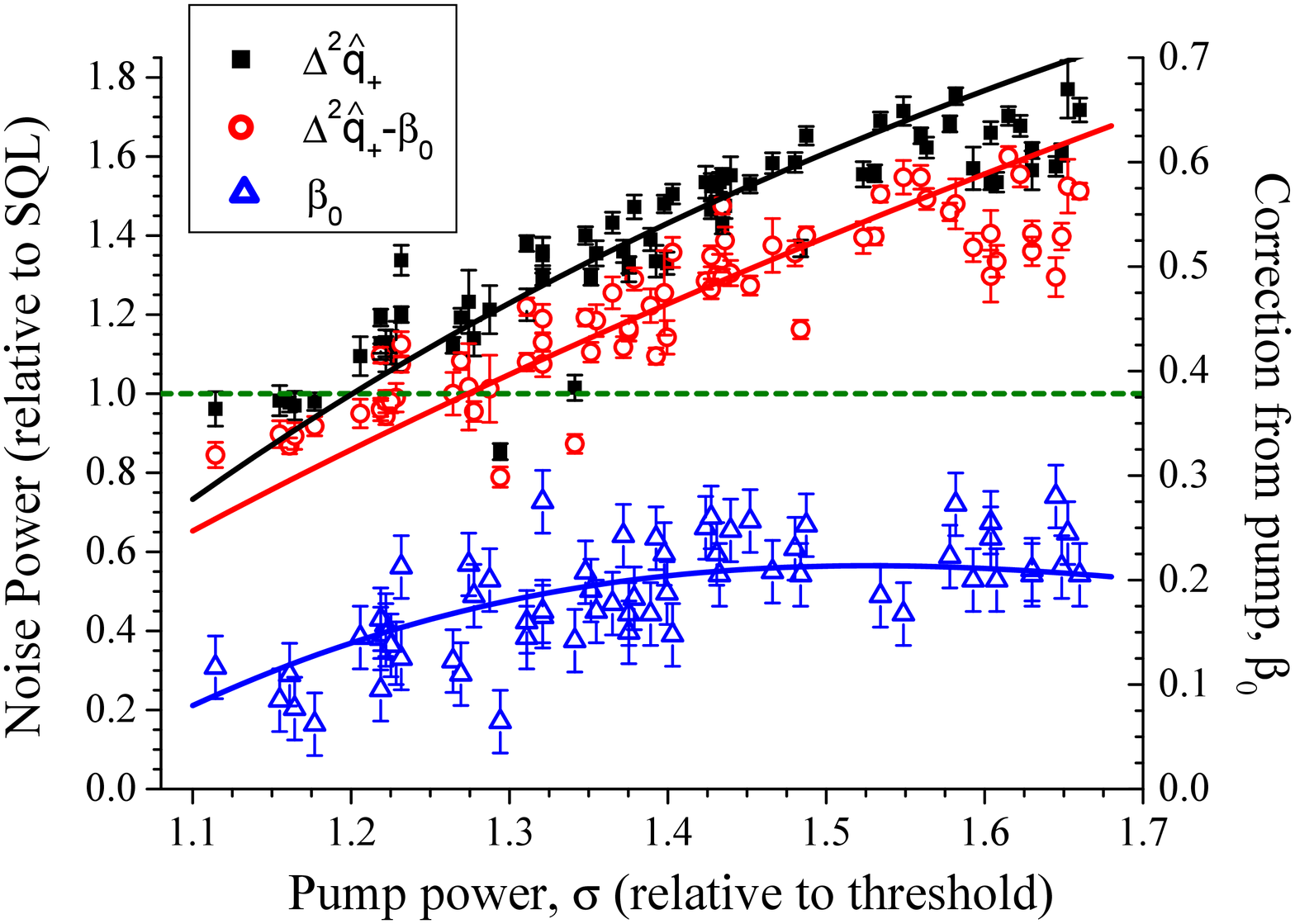,scale=0.31}
\caption{(Color online) Behavior of $\Delta^2\hat{q}_+$, with and without 
correction $\beta_0$ owing to correlations with the pump amplitude, as a function of 
$\sigma$. $\Delta^2\hat{q}_+$: full squares; $\Delta^2\hat{q}_+'$: open 
circles; $\beta_0$: open triangles. Solid lines correspond to physical model 
with pump detuning and excess noise as free parameters.}
\label{fig3}
\end{figure}

Typical noise spectra of sum and difference of twin beam quadratures are presented 
in Fig.~\ref{fig2}, as functions of analysis cavities' detuning relative to 
their bandwidth, $\Delta$. Phase noise $\D^2\hat q_\pm$ is measured for 
$\D=0.5$ and also (partially) for $\D\cong 1.4$. Far off ($|\D|\geq 2.5$) and on 
exact resonance ($\D=0$) the amplitude noise $\D^2\hat p_\pm$ is measured. 
The squeezed difference of amplitude quadratures $\D^2\hat p_-=0.53(2)$ and the 
shot noise limited sum of phase quadratures $\D^2\hat q_+=0.99(2)$ suffice to 
demonstrate bipartite entanglement, since $\D^2\hat p_-+\D^2\hat q_+=1.52(3)<2$.

Quantum correlations between pump and downconverted fields are demonstrated in 
the full line curve. It shows that $\D^2\hat q_+$ can be reduced by using information 
from the pump beam amplitude. When corrected by $\beta_0=0.13(3)$, the {\it shot noise 
limited} sum of phases $\D^2\hat q_+$ becomes {\it squeezed}: $\Delta^2\hat q_+'=0.86(2)$. 
The generalized criterion of Eq.~(\ref{ineqtriple}) assumes the improved value 
$\D^2\hat p_-+\D^2\hat q_+'=1.39(3)<2$.

The behaviors of $\beta_0$, $\D^2\hat q_+$, and $\D^2\hat q_+'$ are 
presented in Fig.~\ref{fig3}, as functions of pump power relative to 
threshold, $\sigma$. Each experimental point was taken from curves 
similar to those of Fig.~\ref{fig2}. The solid lines were calculated 
from the standard linearized OPO theory~\cite{optcomm04}. As described in 
Ref.~\onlinecite{josaboptquinfo}, the theory 
has to be corrected to include extra noise acquired by the intracavity pump field and 
which is not related to the parametric process. We model this by simply adding noise 
to the input pump field. This noise depends on the OPO cavity detuning for the pump 
field~\cite{josaboptquinfo}. Since in our present experiment we do not have 
precise control or knowledge of OPO cavity detunings for pump, $\D_0'$, and downconverted 
fields, $\D'$ (all normalized to the OPO cavity bandwidth for the twin beams), these 
are used as free parameters to fit the data of Fig.~\ref{fig3}. Furthermore, the 
excess phase noise added to the input pump, $S_{q_0}^{\mathrm{in}}$, which is deduced from 
independent measurements of the reflected pump beam for $\D_0'=0$ and $\sigma\approx1$ 
($S_{q_0}^{\mathrm{in}}\approx 23$), also has to be adjusted for the nonzero detunings.

We verify that $\D^2\hat q_+$ is squeezed close to threshold and its noise increases as the 
pump power is increased, crossing the shot noise value at $\sigma\cong 1.2$. Its behavior 
was studied in detail in Ref.~\onlinecite{josaboptquinfo}. The correction 
term $\beta_0$ is always nonzero, varying from $\beta_0\approx 0.10$ to $\beta_0\approx 
0.23$ for increasing $\sigma$. This is in agreement with the theoretical prediction 
of Ref.~\onlinecite{prltrientangopo}, since the degree of triple correlations should be maximum 
close to $\sigma=1.5$, where all fields have approximately the same intensity. In particular, for 
$\sigma\leq 1.3$, the correlation between $\hat p_0$ and $\hat q_+$ reveals or increases the 
squeezing value in $\D^2\hat q_+'$, attesting its quantum nature. Better control of 
the detunings~\cite{pfisterentang} would probably decrease the scattering of the data points. The theoretical 
model is in good agreement with the experiment. The parameters that best fit the data are 
$\D_0'=0.2$, $\D'=0.26$, and $S_{q_0}=15$. From these results, we can surmise that, for $\D_0'=\D'=0$, 
$\hat q_0$ and $\hat q_+$ should be strongly correlated. 

In summary, twin beam entanglement produced by an above-threshold OPO can be improved 
by using the quantum correlations with the pump beam demonstrated here. To our knowledge, 
this is the first experimental demonstration of three-color quantum correlations. They 
can be used, for instance, to increase the fidelity of quantum information distribution. The 
measurement of triple optical quantum correlations is a necessary first step en route to 
the observation of three-color entanglement in the above-threshold OPO.

\section*{Acknowledgments}

This work was supported by Funda\c{c}\~ao de Amparo \`a Pesquisa do Estado de S\~ao Paulo 
(FAPESP) and Conselho Nacional de Desenvolvimento Cient\'\i fico e Tecnol\'ogico (CNPq, through 
{\it Instituto do Mil\^enio de Informa\c c\~ao Qu\^antica}). 


\end{document}